\documentclass[aps,pra,twocolumn,showpacs,amsmath,a4paper,amssymb,10pt]{revtex4-1}
\usepackage{graphicx}
\usepackage{natbib}
\usepackage{epstopdf}
\usepackage{amsmath,amssymb}
\newcommand{\hh}{H$_2$\,}
\newcommand{\hhp}{H$_2^+$\,}

\begin{document}

%\newtheorem{theorem}{Theorem}

%\shorttitle{Channel-resolved subcycle interferences of electron wave packets}
%\shortauthor{X. Xie, S. Roither, D. Kartashov, L. Zhang, A. Baltu\v{s}ka, M. Kitzler}

%\title{subcycle interference of electron wave packets in H$_2$}
\title{Channel-resolved subcycle interferences of electron wave packets emitted from H$_2$ in two-color laser fields}

\author{Xinhua Xie} 
\email[Electronic address: ]{xinhua.xie@tuwien.ac.at}
\author{Stefan Roither}
\author{Daniil Kartashov}
\author{Li Zhang}
\author{Andrius Baltu\v{s}ka}
\author{Markus Kitzler}
\address{Photonics Institute, Vienna University of Technology, Gusshausstrasse 27, A-1040 Vienna, Austria}
%\date{\today}

\begin{abstract}
We report on the observation of subcycle interferences of electron wave packets released during the strong field ionization of \hh with cycle-shaped two-color laser fields.
With a reaction microscope, channel-resolved photoelectron momentum distribution are obtained for different final products originating from single ionization of \hh.
Our results show that the subcycle interference structures of electron wave packet are very sensitive to the cycle-shape of the two-color laser field.
The reason is that the ionization time within an optical cycle is determined by the cycle-shape of the laser field.
The subcycle interference structures can be further used to get the subcycle dynamics of molecules during strong field interaction.
\end{abstract}

\keywords{wave packet interference; strong field ionization; molecular dynamics}

\maketitle

\section{Introduction}

% background

% WHY interesting?
Both electronic and nuclear vibrational dynamics play a crucial role in molecular reactions, such as molecular ionization, dissociation and isomerization \cite{yamanouchi02}.
In general, nuclear vibrational dynamics happens on time scales from tens of femtosecond up to hundreds of femtosecond, while the valence electronic dynamics takes place on the sub-femtosecond/attosecond time scale\cite{Ranitovic21012014}.
Therefore, techniques with attosecond temporal resolution are required to gain insight into the dynamics of valence electrons in molecules.
Experimental techniques such as attosecond extreme-ultraviolet or x-ray transient absorption spectroscopy \cite{Gulielmakis2010,he10,chini12,loh13}, high harmonic spectroscopy \cite{olga09:co2,shiner11,worner11}, and photoelectron spectroscopy based on electron wave packet (EWP) interferences \cite{gopal09,huismans11,xie12,Boguslavskiy2012,zipp14,xie15prl}, have been demonstrated in studies of attosecond electronic dynamics in atoms and molecules \cite{krausz09}.
To retrieve the motion of valence electrons in a molecule, not only attosecond temporal resolution is required but also information on the involved molecular orbitals and the geometry of the molecule is critical.
Since the EWPs released during tunneling ionization of molecules carry phase information on the molecular orbital where they are emitted from \cite{xie07}, this information can be retrieved from the interference patterns of EWPs in photoelectron spectra.
In the past years, EWP interferometry has been applied in studies of ionization dynamics  \cite{xie12}, the molecular orbital \cite{meckel08}, and the influence of the ionic Coulomb potential \cite{huismans11,xie13njp}.
The phase of the interfering EWPs can be reconstructed, and dynamical information taking place during the strong field interaction can be read out from the interference pattern with attosecond temporal resolution \cite{xie12}.

Four types of strong-field-induced EWP interferences in molecules can be distinguished:
The first type are the inter-cycle interferences (ICI) produced by EWPs released during different optical cycles which lead to ATI(above-threshold ionization)-like structures in the momentum or the energy distribution of photonelectrons \cite{arbo10,Boguslavskiy2012}.
The second type are the so-called subcycle interferences (SCI) formed by EWPs detached during different half cycles within one optical cycle \cite{linder05,xie12,arbo06,arbo14}.
The third type of interferences are formed by EWPs removed from the system during the same quarter of one optical cycle due to scattering on the ionic potential \cite{huismans11,spanner04,Bian2012}.
The fourth type are multi-center interferences due to EWPs scattering on different nuclei of a molecule \cite{yurchenko04,spanner04}.
Accurately defined momentum-to-time mapping and a precise identification of the type of observed interference fringes is necessary for retrieving the information of electronic dynamics and structures from the measured interference pattern.
Previous studies reveal that it is important to consider the influence of the ionic Coulomb potential \cite{huismans11,xie12,xie13njp,zhang14_2}.
However, this is not a trivial task.
Besides, because of the mixture of different kind of interferences and complicated low-energy structures (LES) \cite{blaga09,quan09,liu09,Dimitrovski14}, the fringe positions will be modified which affects the precision of the phase reconstruction \cite{arbo10}.

To obtain a well-defined interference pattern of EWPs, a cycle-shaped laser field is beneficial, because the EWP interference pattern are induced by EWPs released at different time. Their release time is this extremely sensitive to the shape of the laser field and can be controlled by the cycle shape of the laser field.
There are several ways to obtain such laser fields.
One of the most straightforward methods is to stabilize the carrier-envelope phase of a few-cycle laser pulse.
Another method is to lock the phase delay between multiple laser fields with different colors.
% what have already done
Such cycle-shaped laser fields have already been applied in studies of controlling high harmonic generation \cite{Goulielmakis08,Haessler14}, molecular orientation \cite{De2009} and dissociation \cite{ray09}, and single and double ionization of atoms \cite{xie12,zhang14_2,zhang14}.

% what will we present in this manuscript? the question to be answered
In this work, we employed a two-color laser field with precise control on the relative phase between a fundamental laser field and its second harmonic.
In a previous study \cite{xie12}, we have demonstrated that subcycle ionization dynamics can be retrieved from the SCI patterns of atomic targets created by such two-color laser fields.
It has been shown that with a cycle-shaped two-color laser field the final momentum of electrons can be shifted to bigger momentum values to avoid overlapping with the complex LES and to minimize influence from the Coulomb potential \cite{xie12}.

In case of molecules, because of small differences in the ionization potentials and different angular dependence of ionization rates, tunnel ionization may happen from lower lying molecular orbitals rather than the HOMO \cite{Wu2012}.
Moreover, the strong field interaction may lead to dissociation of the molecule along different pathways \cite{xie12_2}.
To get access to the electron dynamics of molecular ionization, discrimination between these different pathways is necessary.
Here we extend the application of using subcycle EWP interferometry from atoms to molecules with coincidence measurements which allows disentangling measured the different contributions of pathways to the interference structures.
For our proof-of-principle measurements we chose the simplest molecule, hydrogen (\hh), as the target.

\begin{figure}[htbf]
\centering
\includegraphics[width=0.5\textwidth,angle=0]{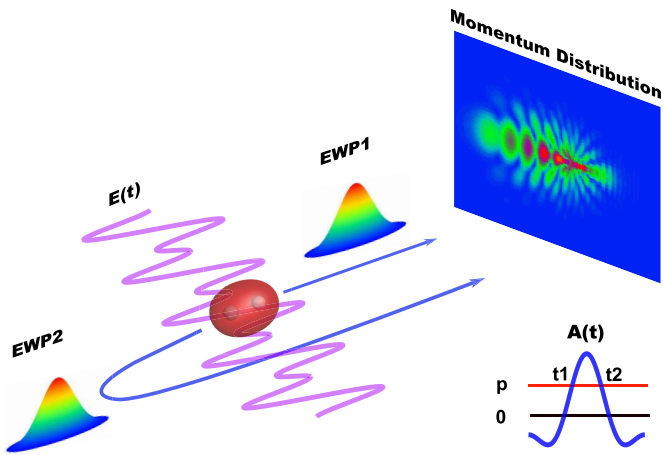}
\caption{A schematic view of subcycle EWP interferences. When a molecule is exposed to a two-color laser field, shown as the magenta line, EWPs will be released around the field's peaks. In a unit cell of the pulse, the EWPs released during each half cycle (t1 and t2) will lead to the same final momentum p, which lead to interference fringes in the momentum space. } \label{fig:peda}
\end{figure}

A schematic view of subcycle EWP interference is shown in Fig.~\ref{fig:peda}. The EWPs released at different times within one laser optical cycle may end at the same final momentum and therefore interfere with each other in the momentum space. Measurement of the photoelectron momentum distribution with a reaction microscope allows recording these interferences \cite{doerner00,ullrich03}.

\section{Experiments}

% How did we do? % advantages
Coincidence measurements of electrons and ions were performed for \hh with a reaction microscope. A laser beam from a home-built Ti:Sapphire laser amplifier system with a center wavelength of 795 nm, a repetition rate of 5 kHz and a pulse duration (full width at half maximum of the intensity) of about 25 fs is superimposed with the second harmonic beam generated by a 500 $\mu$m type-I BBO crystal. The pulse duration of the second harmonic pulse is 46 fs according to self diffraction measurements.
The second harmonic beam is polarized parallel to the fundamental beam and the peak laser intensities are about $6\times 10^{13}$ W/cm$^2$ for each beam.
The group velocity delay between the two laser beams were compensated by calcite plates and a pair of fused silica wedges which was also used to adjust the phase delay between the two colors in steps of $0.06\pi$.
The calibration of the relative phase between the two colors and the laser peak intensities is performed from measurements of helium, as described in Ref.\cite{xie12}.
A weak homogeneous dc field of 2.5 V/cm is applied along the time-of-flight (TOF) spectrometer to accelerate electrons and ions towards two position-sensitive multi-channel plate detectors.
Additionally, a homogeneous magnetic field of 6.4 gauss ensures 4$\pi$ detection of electrons.
The beam of hydrogen molecules with a diameter of about 170 $\mu$m is prepared by supersonic expansion through a nozzle with a diameter of 30 $\mu$m and collimated with a two-stage skimmer before the ultra-high vacuum interaction chamber (about 1.3$\times 10^{-10}$mbar).
TOFs and positions of electrons and ions are recorded and the momentum vectors of all particles are retrieved in the off-line data analysis.
Momentum conservation condition between electrons and ions can be applied to achieve coincidence selection to minimize the background signals.
To ensure a high efficiency of coincidence detection, the ionization rate is kept at about 0.4 ionization events per laser shot.
More details on the experimental setup can be found in our previous publications \cite{xie12,xie12_2}.

\section{Results and discussion} % what did we find? why?

\subsection{\hh in two-color laser fields}

\begin{figure}[ht]
\centering
\includegraphics[width=0.5\textwidth,angle=0]{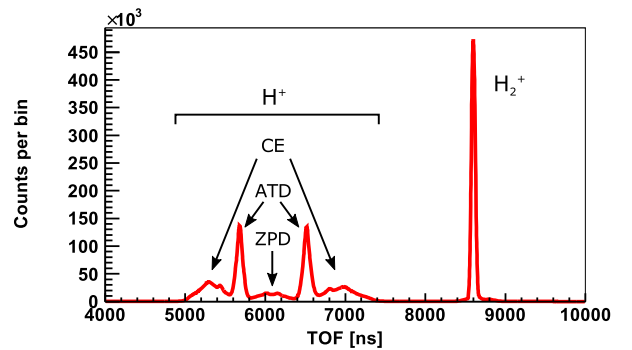}
\caption{ Time-of-flight spectrum of \hh interacting with two-color laser fields. In the H$^+$ distribution, there are three regions: zero-photon-dissociation (ZPD), above-threshold-dissociation (ATD) and Coulomb explosion (CE).} \label{fig:tof}
\end{figure}
When a hydrogen molecule interacts with a strong laser field, one electron can be removed through tunnel ionization.
After single ionization, the molecule will reach a cationic state.
H$_2^+$ may then dissociate into a proton and a hydrogen atom, or H$_2^+$ can be further ionized and eventually lead to Coulomb explosion into two protons.
A typical time-of-flight spectrum of H$_2$ in a strong two-color laser field is presented in Fig.~\ref{fig:tof}.
The single ionization (\hhp), dissociation ($H^++H$) and Coulomb explosion ($H^++H^+$) pathways can be well distinguished.
As marked in Fig.~\ref{fig:tof}, there are two pathways for the dissociation of \hhp: above-threshold dissociation (ATD) leading to protons with high energy, and zero-photon dissociation dissociation (ZPD) leading to low-energy protons \cite{moser2009ultraslow,posthumus2000}.

\begin{figure}[ht]
\centering
\includegraphics[width=0.5\textwidth,angle=0]{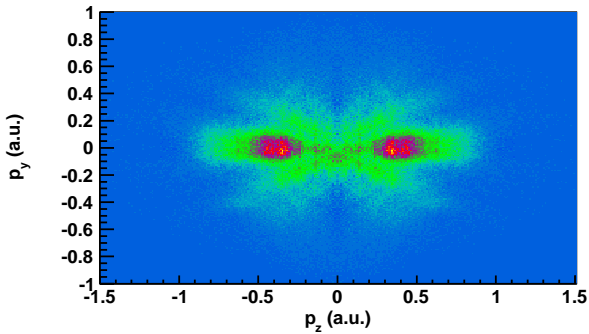}
\caption{ A slice through a measured electron momentum distribution in the $x-z$ plane (laser field polarization along $z$-direction) with a condition $|p_{y}|<0.1 a.u.$ and integration over all relative phases. Momentum conservation conditions between one electron and H$_2^+$ is applied to the measured data to ensure coincidence selection. To enhance the visibility of structures induced by electron wave packet interferences, a gaussian function is subtracted for each relative phase. } \label{fig:pxz}
\end{figure}
First we focus on the single ionization leading to H$_2^+$.
A measured electron momentum distribution correlated with H$_2^+$ is shown in Fig.~\ref{fig:pxz}, integrated over all relative two-color phases.
In the distribution there are two clear types of structures: finger-like patterns due to scattering of EWPs on the parent nucleus and ATI-like ring structures.
Here in this paper, we focus on SCI of EWPs released during strong field interaction.
However, in Fig.~\ref{fig:pxz}, there are no clear structures of subcycle EWP interference.
The reason is that the subcycle EWP interference is very sensitive to the cycle shape of the laser field.
As the electron momentum distribution in Fig.~\ref{fig:pxz} is integrated over all relative phases between the two color fields, the structure of SCI is smeared out.

\subsection{Photoelectron momentum distribution over relative phase}

\begin{figure}[ht]
\centering
\includegraphics[width=0.5\textwidth,angle=0]{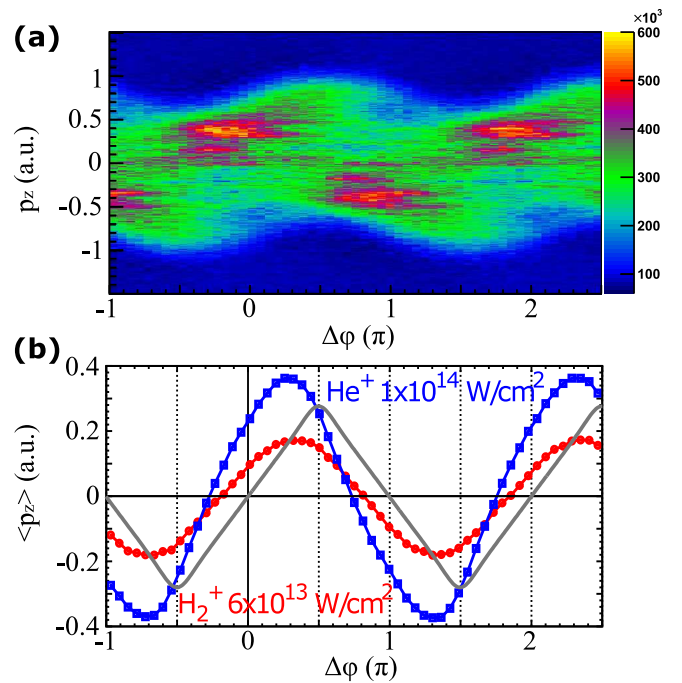}
\caption{ (a) Measured ion momentum distribution along the laser polarization direction over the relative phase between the two colors. To enhance the visibility of structures induced by electron wave packet interferences, a gaussian function is subtracted for each relative phase. (b) The mean value of the momentum distribution along the laser polarization direction as a function of the relative phase between the two colors. The gray line represents the simulated results using SFA. } \label{fig:pz_delay}
\end{figure}

Measured distributions of the H$_2^+$ ionic momentum along the laser polarization direction ($p_z$) over the relative phase are shown in Fig.~\ref{fig:pz_delay}(a).
Clear 2$\pi$-periodic modulations can be seen.
The mean momentum value oscillates over the relative phase [depicted in Fig.~\ref{fig:pz_delay}(b) as red circles] and reaches maximum offset at $\Delta\phi=0.35+n\pi, n\in \mathbb{Z}$.
The mean value of the momentum distribution is determined by the shape of the laser field vector potential.
For relative phase 0, the vector potential of the two-color field is symmetric, which according to prediction of the simple-man's model within the strong field approximation (SFA) \cite{Corkum94} should lead to zero mean value.
As shown previously, the observed offset from 0 is due to the ionic Coulomb potential \cite{bandrauk04,bandrauk05,xie13njp}.
The phase shift due the Coulomb potential is about -0.2$\pi$ as compared with the simple-man's results, in which the Coulomb effect is neglected \cite{bandrauk04,bandrauk05}.
The momentum mean value of a measurement on helium with higher laser peak intensity ($1\times 10^{13}$ W/cm$^2$ for each color) \cite{xie12} is plotted in Fig.~\ref{fig:pz_delay}(b) as blue squares.
The amplitude of the mean value oscillation for helium is almost twice as that for hydrogen because of the higher peak laser intensity.
On the other hand, we notice that the phase shift of the helium measurements is about -0.3$\pi$, i.e.  more than that of the hydrogen measurement.
This contradicts our previous finding that the Coulomb effect is stronger for lower laser peak intensity \cite{xie13njp}.
The reason may be the participation of excited states or non-adiabatic effects in the ionization process \cite{Yudin2001,Spanner2011}.

\subsection{Subcycle interference of electron wave packet}

\begin{figure*}[htbp]
\centering
\includegraphics[width=0.75\textheight,angle=0]{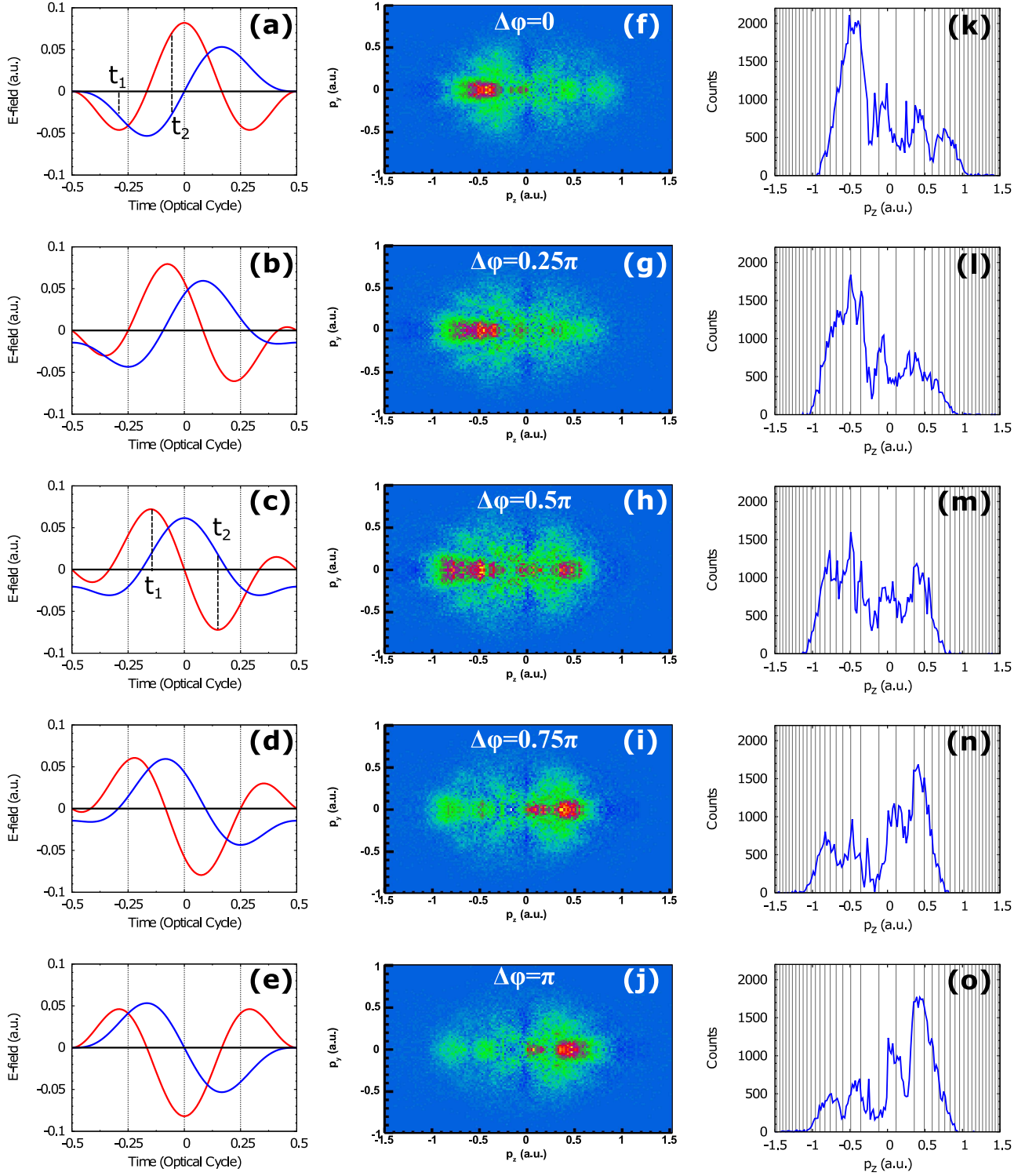}
\caption{ (a-e) Electric fields (red lines) and vector potentials (blue lines) for relative phases 0, 0.25$\pi$, 0.5$\pi$, 0.75$\pi$ and $\pi$. (f-j) Measured electron momentum distributions in the laser polarization plane with subtraction of a gaussian function for the five relative phases. (k-o) Momentum distributions along the laser polarization direction with $|p_{x,y}|<0.1$ a.u. for the five relative phases. Vertical gray lines indicate the positions of the ATI peaks. } \label{fig:ddf}
\end{figure*}
To observe SCI structures we investigate electron momentum distributions for certain relative phases between the two colors, since for electrons the momentum resolution is much higher along the directions perpendicular to the laser polarization direction that for ions.
The measured electron momentum distribution in the laser polarization plane is illustrated in Fig.~\ref{fig:ddf}(f-j) for five different relative phases (0, 0.25$\pi$, 0.5$\pi$, 0.75$\pi$ and $\pi$).
The structures in the momentum distribution look similar to those measured for helium \cite{xie12}.
There are sharp ATI-like peaks in the low momentum region as shown in the $p_z$ distribution [Fig.~\ref{fig:ddf}(k-o)] and finger-like structures in the 2D momentum distribution [Fig.~\ref{fig:ddf}(f-j)].
SCI structures can been seen in the 1D momentum distribution as big humps [Fig.~\ref{fig:ddf}(k-o)].
However, in details the distributions are different from those observed for helium because of different laser intensities, different ionic potential and energy structure.

First, we focus on the results with relative phase 0.
According to the shape of the vector potential [Fig.\ref{fig:ddf}(a)], the momentum distribution should be symmetric along $z$-direction.
However, due to the Coulomb effect, the EWP released before the peak will be driven back and scatter with the parent ion which leads to the appearance of clear finger-like holographic structures.
In the results for helium, there are no obvious SCI structures.
The reason is that for helium the ionization mainly happens near the major peak within one optical cycle and therefore the SCI of EWPs is suppressed.
In contrast, for \hhp there appear SCI fringes for $p_z>0.2$a.u. in the momentum distribution [Fig.~\ref{fig:ddf}(f,k)].
Such a structure within the SFA is induced by the interference of EWPs released at $t_1$, the minor peak within one optical cycle, and $t_2$, before the major peak.
EWPs released at $t_1$  and $t_2$ end at the same final momentum and lead to SCI.
The reason for the more pronounced SCI structure for a relative phase of 0 in hydrogen than in helium may be the following: The ionization potential of helium, 24.6 eV, is higher than that of hydrogen, 15.5 eV.
Therefore, the ionization of helium dominantly happens near the main peak within a laser optical cycle, while in the ionization of hydrogen the two minor peaks may have also contribute considerably.

For a relative phase of 0.5$\pi$, the laser electric field is symmetric [Fig.~\ref{fig:ddf}(c)].
Therefore, within one optical cycle, ionization may happens equally at the two main peaks ($t_1$ and $t_2$) and the released EWPs will interfere with each other in the momentum space because they will end at the same final momentum.
In Ref.~\cite{xie12}, the SCI with observed at a relative phase 0.5$\pi$ is applied to retrieve the phase of the released EWPs and to investigate the electron dynamics during strong field ionization.
As shown in Fig.~\ref{fig:ddf}(h), there are clear SCI structure on the negative momentum side.

When the relative phase between the two colors varies, the cycle structure of the laser field varies accordingly.
Since tunneling ionization is very sensitive to the field strength of a laser field.
Therefore, the ionization time within an optical cycle can be well controlled with the laser cycle shape and leads to the control of SCI.
As presented in Fig.~\ref{fig:ddf}, it is clearly shown that the interference structures, especially the SCI structures, strongly depend on the relative phase of the two colors.

\subsection{Subcycle interference of electron wave packet for dissociation pathways}

After single ionization in a two-color field, \hhp may dissociate into a proton and a hydrogen atom through ATD or ZPD.
We can distinguish the dissociation pathways based on the kinetic energy release during the dissociation process.
By coincidence gating we can obtain the corresponding photoelectron spectra.
From the channel-resolved photoelectron spectra we can get access to the electron and nuclear dynamics which leads to the certain dissociation channels.
For \hh the different channels can be easily separated by the proton momentum.
We select the proton momentum in the range of $6<\lvert p_z \rvert <11$ a.u. for ATD, and $\lvert p_z \rvert<6$ a.u. for ZDP.

\begin{figure}[ht]
\centering
\includegraphics[width=0.5\textwidth,angle=0]{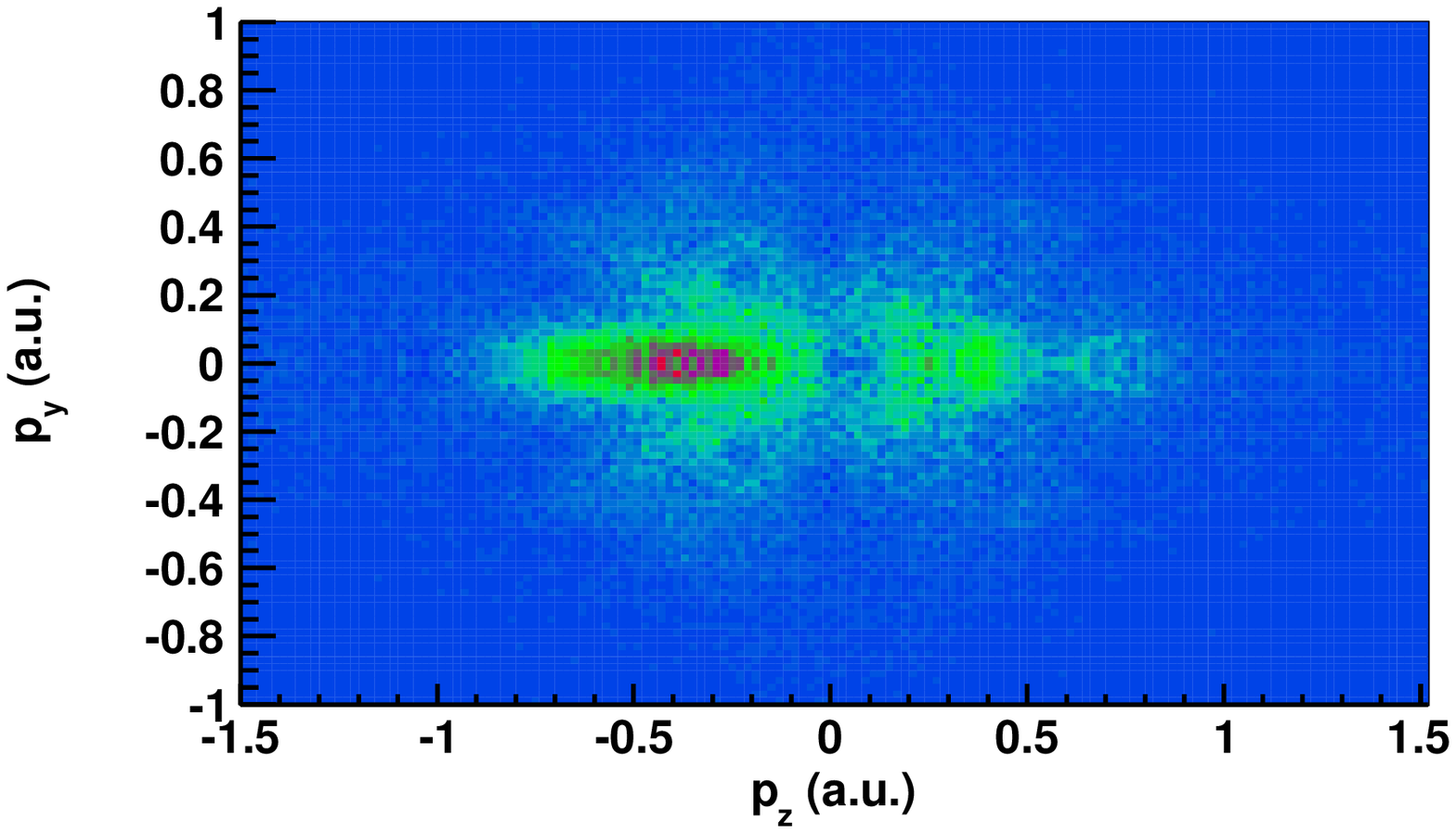}
\caption{ Electron momentum distribution for a relative phase of 0 for the ATD pathway. } \label{fig:diss}
\end{figure}

In Fig.~\ref{fig:diss} electron momentum distributions along the laser polarization direction are plotted for the ATD pathway and a relative phase of 0.
The structure is similar to that of the \hhp pathway [Fig.~\ref{fig:ddf}(f)].
Due to the limited momentum resolution, further comparison in detail is not meaningful for this measurement.
Our experimental observation of SCI in the strong field ionization of \hh constitutes a proof-of-principle experiment of channel-resolved EWP interferometry based on SCI.
Similarly to channel-resolve ATI spectroscopy based on the ICI of EWPs \cite{Boguslavskiy2012}, channel-resolved subcycle EWP interferometry can be applied to investigate the relation between electron dynamics and nuclear dynamics in molecules induced by strong field interaction.

\section{Conclusion}

% conclusion and outlook
In conclusion, we demonstrated a proof-of-principle experiment of channel-resolved EWP interferometry based on SCI of EWP released during strong field interaction.
We report on the observation of SCI in the strong field interaction of hydrogen molecules with cycle-shaped two-color laser fields.
It is found that the structure of SCI is very sensitive to the cycle shape of the two-color laser fields.
Because in the measurement ATD and ZPD of \hhp can be distinguished with proton energy, channel-resolved electron momentum spectra could be obtained for the ionization and dissociation channel for \hhp.
The channel-resolved subcycle EWP interferometry demonstrated here can be employed for studies of multi-electron and multi-orbital effects in the laser field-induced ionization and dissociation of molecules \cite{Boguslavskiy2012,Xie2014prx,Xie2014prl}.

\section{Acknowledgement}

This research was financed by the Austrian Science Fund (FWF) under grants P25615-N27, P28475-N27, P21463-N22, P27491-N27, and SFB-F49 NEXTlite, and by a starting grant from the European Research Council (ERC project CyFi).

\end{document}